\documentclass[twocolumn,showpacs,prl,superscriptaddress]{revtex4}
\usepackage{subfigure}
\usepackage{amsmath}
\usepackage{graphics}
\usepackage{rotating}

\newcommand{\br}{{\bf r}}

\newcommand{\beqa}{\begin{eqnarray}}
\newcommand{\eeqa}{\end{eqnarray}}

\begin{document}

\title{ Wigner crystallization in  a single and bilayer graphene }
\author{Hari P. Dahal}
\affiliation{Department of Physics, Boston College, Chestnut Hill,
 MA, 02467}
 \affiliation{Theoretical Division, Los Alamos National Laboratory,
Los Alamos, New Maxico 87545}

\author{Tim O. Wehling}
\affiliation{Theoretical Division, Los Alamos National Laboratory,
Los Alamos, New Maxico 87545}
 \affiliation{I. Institut f\"{u}r
Theoretische Physik, Universit\"{a}t Hamburg, Jungiusstra\ss e 9,
D-20355 Hamburg, Germany}

 \author{Kevin S. Bedell}
\affiliation{Department of Physics, Boston College, Chestnut Hill,
    MA, 02467}

\author{Jian-Xin Zhu}
\affiliation{Theoretical Division, Los Alamos National Laboratory,
Los Alamos, New Maxico 87545}

\author{A. V.  Balatsky}
\affiliation{Theoretical Division, Los Alamos National Laboratory,
Los Alamos, New Maxico 87545}
\affiliation{Center for Integrated
Nanotechnology, Los Alamos National Laboratory, Los Alamos, New
Maxico 87545} \email[] { avb@lanl.gov, http://theory.lanl.gov}

\begin{abstract}
We study the possibility of Wigner crystallization  in both
single- and and bi-layer graphene using a real space tight binding
model. In addition to verifying our earlier prediction for single
layer graphene, we predict that the bilayer graphene can undergo a
Wigner crystal transition at low enough carrier density thus
offering a possibility of tunable and bipolar Wigner crystal. We
find that aside from the Coulomb interaction of the carriers of
two layers that stabilizes the charge ordered state, so does the
inter layer coupling.
\end{abstract}

\maketitle

The study of low dimensional system covers a great deal of area in
the research field of physics. In the past couple of years, the
experimental success of fabricating single \cite{novoselov2004,
novoselov2005} and bi-layer graphene~\cite{novoselov2006} has
enriched the interest of studying the properties of low
dimensional strongly interacting electronic systems both
theoretically and experimentally. The low dimensional system of
graphene is more interesting because many of the properties of
these systems are quite different from that of the conventional
two dimensional system; such as low temperature
\cite{novoselovnature2005,novoselov2006} and room temperature
\cite{novoselovscience2007} quantum Hall effect, and suppression
of weak localization of the carriers \cite{morozov2006,
gorbachev2007}. From technological point of view, theses systems
open up possibilities of  atomic-scale electronic devices which
can go beyond the  limitations of the silicon-based electronics.
In addition to the small size of these materials, the very high
mobility of the carriers of the single and bilayer graphene looks
to be promising for devices which depend up on the transport
properties of the carriers.

The importance of studying the single- and bi-layer graphene not
only comes from the fact that they have different physical
properties compared to the other conventional system such as two
dimensional electron gas (2DEG) prepared in the heterostructure of
GaAs and InGaAs but also from the fact that their properties are
very different between themselves. One of the best example is the
experimental verification of the different quantum Hall effect in
the single \cite{novoselovnature2005} and bilayer graphene
\cite{novoselov2006}. Both single and bilayer system show weak
localization effect but with different minimum zero-bias
conductivity. The theoretical prediction of the Klein paradox in
the single layer graphene is another example
\cite{katsnelson2006}.

Structurally, graphene has a hexagonal lattice structure having
two inequivalent lattice sites, A and B, per unit cell. Bilayer
graphene is obtained from two identical single layer graphene with
a $z$-axis stacking in a specific way known as Bernal stacking
\cite{nilsson2006}. The intra layer nearest-neighbor hopping
energy is about ten times stronger than the inter layer hopping.
So, to the lowest order approximation, each layer of the bilayer
graphene is almost similar to the single layer graphene, except
that the two layers are coupled via small inter layer hopping. The
effective difference can be seen in the low energy dispersion
relation of the charge carriers in the conduction band which for
the single and bilayer graphene can be written as, $\varepsilon_k
= \pm \hbar v_f k$, and $E_k = \pm t_\perp \pm \sqrt{t_\perp^2 +
\varepsilon_k^2}$ ,\cite{nilsson2006} respectively, where, we set
$\hbar =1$. These dispersion relations are obtained using tight
binding model where $v_F = \frac{3ta}{2}=5.8eV \dot{A}$ is the
Fermi velocity, $t=3.2eV$ is the intra layer nearest-neighbor
hopping energy, and $t_\perp = 0.1t$ is the inter layer hopping
energy. For bilayer graphene for small $k$, the dispersion
relation for two of the branches can be approximated as, $E_k =
\pm \frac{k^2}{2m} $. So due to the inter layer hopping the
massless particles of the single layer graphene become massive
where the mass of the carriers, $m$, is related to inter-layer
hopping energy by, $m = t_\bot/v_{F}^{2}$. An important point to
note here is that the dispersion relation of the single layer
graphene is quite different than that of the 2DEG system whereas
the dispersion relation of the bilayer graphene has similar form,
i.e., parabolic dispersion, as that of 2DEG system. Flat
dispersion in bilayer effectively lowers the kinetic energy for
particles with small momenta. This effect will become important
for our discussion of Wigner Crystallization (WC).

Here we want to study the charge ordering effects in single versus
bilayer graphene. Difference in electronic spectra could manifest
itself as a difference in the possible charge ordered states.
Indeed we find this to be the case. We find that WC can occur in a
bilayer system at small doping limit in zero magnetic field. On
the other hand the single layer graphene is stable against charge
ordering \cite{dahal2006}. WC is a physical phenomenon envisioned
by Wigner \cite{wigner1934}, in which the crystallized phase
appears when electrons localize and form a crystal to minimize the
potential energy, while paying the concomitant kinetic energy cost
which arises from localization. This phase arises in 2DEG system
which has parabolic dispersion relation as we decrease the carrier
density, $n$, of the system because the ratio of the potential to
kinetic energy varies as $\frac{1}{\sqrt{n}}$ which can be made as
large as it is needed to get the crystallized phase by making the
system more and more dilute.

This discussion brings  obvious questions about the WC in
graphene: a) Can a single layer graphene have a Wigner crystal
phase even if it has linear dispersion relation? b) Is it obvious
that the bilayer graphene will have Wigner crystal phase since it
has parabolic dispersion realation? c) If bilayer can have
crystallized phase what is the structure of the crystallized
phase? We will answer these questions by calculating the
occupation $\langle n \rangle$, of carriers on lattice sites of
single and bilayer graphene using a real space tight binding
model. If $\langle n\rangle$ is equal for all sites, we will
define the state as homogeneous phase, and if it not equal for all
sites, we define the state as some sorts of inhomogeneous phase,
one of which is the Wigner crystal phase. We will determine the
structure of the inhomogeneous phase with varying $\langle
n\rangle$.

\begin {figure}
\includegraphics[width= 5.0 cm, height = 4.0 cm]{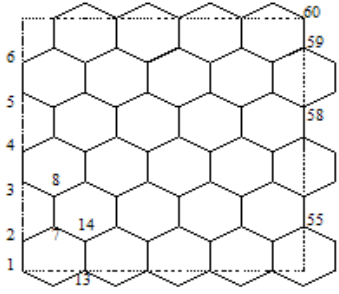}
\caption{Lattice structure of graphene single layer having 60
lattice sites that we used to solve the hamiltonian. The index
number of some of the lattice sites are shown.}
\end{figure}

We now use a real space tight-binding model a) to verify that
single layer graphene remains in the liquid phase, and b) to
predict that bilayer graphene can undergo a crystallization
transition. We study the electronic distribution on the lattice of
single layer and bilayer graphene having 60 atomic sites in each
layer. The real space distribution of the system of 60 lattice
sites that we have chosen and some of the corresponding indices is
shown in Fig. 1. In addition to this layer, the bilayer graphene
has another layer below or above this layer with a shift of the
whole layer by one lattice spacing in the lateral direction. For
the second layer we index the lattice sites with indices 61 to
120. We use the real space tight binding hamiltonian.

The tight-binding Hamiltonian to be solved is as follows:
\begin{equation}
\begin{split}
H = -\sum_{k=l,i,j} t_{ki,lj} c_{ki}^\dagger c_{lj} - \sum_{k\neq
l,i,j}
t_{\bot Aki, Alj} c_{Aki}^\dagger c_{Alj}  \\
+ \frac{1}{2}\sum_{ki\neq lj} V_{ki,lj}n_{ki} n_{lj} + \sum_i V_0
n_{ki\uparrow}n_{ki\downarrow} - \mu\sum_{ki} n_{ki},
\end{split}
\end{equation}
where $i,j$ denote the lattice sites in both single and bilayer,
and $k,l=1,2$ represent index for the first or second layer of
bilayer graphene. For single layer, $k=l=1$. Since the first term
of Hamiltonian has summation over $i,j$ only for $k=l$,
$t_{ki,lj}$ represents the hopping energy of the electrons only
within a layer. We do calculation including only the nearest
neighbor hopping with energy equal to $ t_{ki,lj} = t = 3.2eV$ for
all lattice sites. For the second term of the Hamiltonian the
summation is only for $k \neq l$ which implies that for single
layer graphene this term is zero and for bilayer graphene $t_{\bot
Aki, Alj}$ represents the inter layer hopping energy. In this term
$A$ denotes one of the non-equivalent lattice sites. Since the
bilayer graphene can be obtained by the special stacking called
the Bernal stacking of two single layers grpahene,  electrons can
hop between two layers only through one of the sublattices which
we have chosen to be A. For the summation of this term, since one
of the two layers is shifted laterally by one lattice spacing, if
$i$ is an even numbered lattice index, $j$ will be an odd numbered
lattice index or vice-versa. $V_0$ is the on site coulomb
repulsion, $V_{ki,lj}$ is the interacting potential between two
electrons at position $r_{ki}$ and $r_{lj}$, $\mu$ is the chemical
potential, and $n_{ki}, n_{lj}$ are the occupation number of the
electrons on lattice sites $i, j$ of the layer $k,l$ respectively.
Except otherwise indicated, we use $V_{ki,lj} = \frac{e^2
\exp(-|r_{ki}-r_{lj}|/2a)}{\epsilon(a + |r_{ki}-r_{lj}|)}$, where
$2a$ is the screening length, so that we could regularize the
on-site potential $V_0= V_{ki=lj} =\frac{e^2}{\epsilon a}$ to be
finite. There are indications that Coulomb terms are decaying
faster than $1/r$ that would also be consistent with our
approximation of taking into account only a few lattice sites
\cite{katsnelsonprb2006} \cite{note1}.

We use mean-field theory to simplify the Coulomb interaction term.
The Hamiltonian then can be expressed as,
\begin{equation}
\begin{split}
H_{MF}= -\sum_{k=l,i,j} t_{ki,lj} c_{ki}^\dagger c_{lj} -
\sum_{k\neq l,i,j} t_{\bot Aki, Alj} c_{Aki}^\dagger c_{Alj} \\ +
\sum_i (W_{ki} + U_{ki} - \mu) n_{ki} - C,
\end{split}
\end{equation}
where $W_{ki} = \sum_{lj} V_{ki,lj}\langle n_{lj} \rangle $,
$U_{ki} = \frac{V_0 \langle n_{ki} \rangle}{2} $, $C = \frac{1}{2}
\sum_{ki\neq lj} V_{ki,lj}\langle n_{ki} \rangle \langle n_{lj}
\rangle$ + $\sum_{ki}V_0 \langle n_{ki\uparrow} \rangle \langle
n_{ki\downarrow} \rangle$  is a constant.

\begin {figure}
\includegraphics[width= 8.0 cm, height = 3.50 cm]{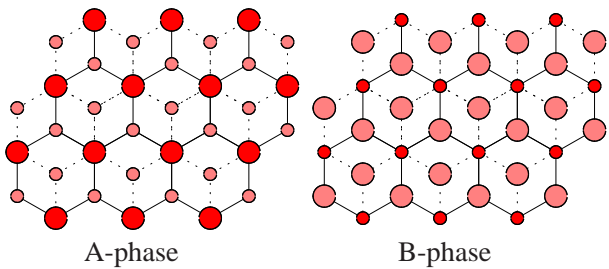}
\caption{ (Color online) Charge distribution in ordered state of
bilayer graphene is shown. Larger circles correspond to larger
charge density on lattice sites. The sites connected by solid
lines belong to the top layer and by dotted lines belong to bottom
layer. The left figure, which we define as A-phase, has charge
ordering where larger charge densities are sitting on sub-lattice
A. The right figure, which we define as B-phase, has charge
ordering where large charge densities are sitting on sub-lattice
B.} \label{FIG:chargeorder}
\end{figure}

Using periodic boundary condition we write down a Hamiltonian
matrix corresponding to,
\begin{equation}
h_{ki,lj}=-t_{ki,lj} - t_{\bot Aki, Alj} + (W_{ki} + U_{ki} -
\mu)\delta_{ki,lj},
\end{equation}
for single and bilayer graphene. For single layer graphene, we
calculate $W_{ki}$ including carriers up to next-next nearest
neighbor. For $60$ lattice sites in a single layer, the
Hamiltonian becomes a $60\times 60$ symmetric matrix. For the
bilayer graphene also, we calculate the contribution to $W_{ki}$
coming from both the layers $k,l$ including the next-next nearest
neighbor. The Hamiltonian for the bilayer graphene becomes a
$120\times 120$ symmetric matrix.

To find the distribution of electrons on the lattice we start with
a random distribution of $\langle n_{ki}\rangle$. Using random
initial input of $\langle n_{ki} \rangle$, we construct
$h_{ki,lj}$ and find the eigenvalues $E_n$ and eigenvectors
$\phi_{ki}^n$. We calculate new $\langle n_{ki}\rangle $ using
following relation,
\begin{equation}
\langle n_{ki}\rangle = \sum_n |\phi_{ki}^n|^2 f(E_n),
\end{equation}
where, $f(E_n) = 1/(e^{\frac{E_n}{k_BT}}+1)$ is the Fermi
distribution function. Using the standard procedure of the mixing
of the old and new $\langle n_{ki} \rangle$, we repeat the
calculation until we reach the convergence of $\langle n_{ki}
\rangle $. We check that the results we obtain do not depend on
initial distribution of $n_{ki}$.

a) {\em Single layer case}. We calculate  occupation number,
$\langle n_i \rangle$, of electrons on each lattice site of the
single layer graphene. We choose different values of the chemical
potential and Coulomb potential to vary the average occupation
number below half filling and evaluate $\langle n_i \rangle$ for
all $i$. We always find $n_i = n_{j}$ for all sites. This defines
a homogeneous charge distribution also known as the liquid phase.
Thus we verify our earlier prediction that the single layer
graphene always remains in the liquid phase. Here we remark that
this homogeneous phase assumes that graphene sheet has no ripples
\cite{Geim07}.

b) {\em Bilayer case}. Similar to the case of single layer
graphene, we calculate $\langle n_i \rangle$ for each lattice
site. In bilayer case, in addition to the Coulomb energy and
chemical potential, we can tune the inter layer hopping energy,
$t_\bot$. First we imagine turning off the Coulomb energy and
study the charge distribution as a function of $t_\bot$. We find
that $\langle n_i \rangle = \langle n_j \rangle$ for all $j=i+2$,
and $\langle n_i \rangle \neq \langle n_j \rangle$ for all
$j=i+1$. This implies that all A-sites have equal occupation.
Similarly, all B-sites have equal occupation. But, site A and B do
not have equal occupation. This defines a commensurate triangular
inhomogeneous phase. Since this inhomogeneity is driven only by
the inter-layer hopping energy, we define this phase as kinetic
energy driven (KED) inhomogeneous phase. Depending up on the
average occupation of the charge carriers on the lattice, also
known as filling factor, we find two distinct inhomogeneous
phases, KED-A and B phase. A schematic representation of the
charge distribution in these two phases is shown in Fig.
\ref{FIG:chargeorder}. We find KED-A phase when the filling factor
is less than half filling and the KED-B phase when it is more than
half filling. We find a cross over between these phases at half
filling. So exactly at half filling the system will be in liquid
phase.

When we turn off $t_\perp$ and turn on Coulomb energy, we again
find the inhomogeneity in the charge distribution. The charge
distribution resembles to that of the B-phase as shown in Fig.
\ref{FIG:chargeorder}. Since this phase is driven by the
inter-layer Coulomb interaction, we define this phase as a Coulomb
interaction driven Wigner crystal (WC) phase. The pattern of the
inhomogeneous charge distribution does not change below and above
the half filling case; of course the system  regains the liquid
phase far above the half filling.

If we include both the inter layer Coulomb interaction and the
inter layer hopping, we find competing phases below the half
filling and cooperating phases above it. Above the half filling
KED-B and WC-B phases cooperate with each other. Below the half
filling, KED-A and WC-B phases compete with each other. Since this
competing phase is more interesting, we are going to explore it
more for a system which has $40\%$ filling.

\begin {figure}
\includegraphics[width= 8.0 cm, height = 6.0 cm]{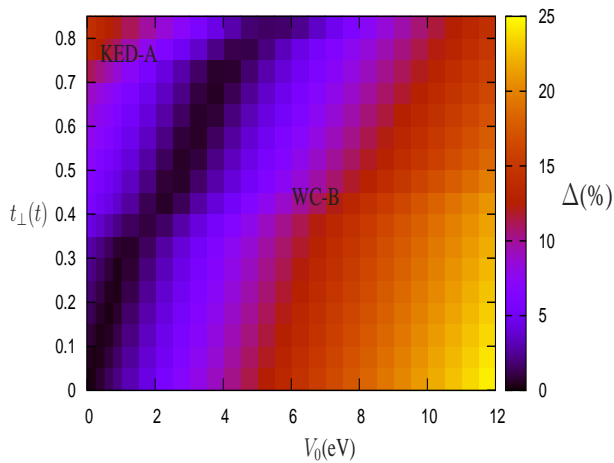}
\caption{(Color online) The order parameter of competing KED-A and
WC-B phases as functions of $t_\bot$ and $V_0$ at $\langle n
\rangle =0.4$. Increasing $V_0$ increases the order parameter of
WC-B phase, whereas increasing $t_\bot$ increases the order
parameter of KED-A phase. The black-colored region represents the
phase boundary between theses two competing phases.}
 \label{competingphase}
\end{figure}

To define the degree of the inhomogeneity of both the A and B
phases we define a charge order parameter, $\Delta =
\frac{2(n_B-n_A)}{n_B + n_A}$. In principle this order parameter
is position dependent, but in our calculation it is
transactionally invariant. For B-phase it gives a positive number
and for the A-phase it gives a negative number. We will be using
the absolute value. The result of the calculation of this order
parameter as functions of $t_\perp$ and $V_0$ is shown in Fig.
\ref{competingphase}.

For a finite $V_0$, an increase in $t_\bot$ decreases the order
parameter of WC-B phase. A further increase in $t_\bot$ destroys
the WC-B phase turning it in to a liquid phase and finally drives
the system to a KED-A phase. Similarly, for a finite $t_\bot$ for
zero $V_0$ the system is in the KED-A phase. Increase in $V_0$
destroys the KED-A phase turning it to a liquid phase and finally
drives the system in to WC-B phase. The black colored region on
the figure separates the two competing phases. For realistic
parameters we expect $V_0 \gg t_{\perp}$ and WC-B state always
wins. In practice disorder and fluctuations will produce mixture
of these two states in real systems.

\begin {figure}
\includegraphics[width= 8.60 cm, height = 6.0 cm]{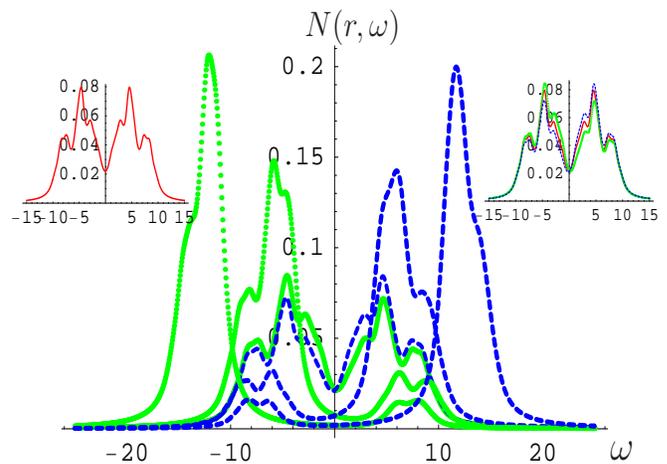}
\caption{LDOS of WC phase as a function of energy for various
$\Delta = 0.19, 1.25, 1.75$ and fixed $t_\perp = 0.1t$. LDOS of
sub-lattice A and B in WC-phase is shown in blue (dahsed) and
green (dotted) curves, respectively. LDOS of a liquid state in a
single layer graphene is shown in Red color (solid curve). One can
clearly see the linear dependence of DOS on energy that is
characteristic of graphene and should be observed for all LDOS in
a liquid. The offset on the Y-axis is due to the effect of the
Lorentian broadening (half-width = $0.25t$) used to make LDOS
curve smooth. LDOS for a WC with $\Delta = 0.19$ is shown in the
upper right figure. If we zoom-in the figure we can see the
emergence of the energy gap in the spectrum (that is not fully
developed due to finite size effects in our calculations). One
also can observe the staggering of the LDOS as a function of
energy for nearest neighbor sites. The site with larger charge
density will have a larger integrated weight on the negative bias
side and vice versa. For $\Delta  = 1.75$ we find fully developed
gap in the spectrum (main panal, highest peak) with the same
staggering between sublattices in WC. The higher values of
$\Delta$ is obtained by taking larger screening length of the
Coulomb potential.}
 \label{FIG:LDOS}
\end{figure}

How one can  detect WC in a bilayer? One experiment that would be
sensitive to the charge ordering at atomic scale is scanning
tunneling spectroscopy. It would allow one to measure spatial
variations of local density of states. To better characterize WC
states we calculate the local density of states (LDOS)
$N_k(\br_i,\omega) = \sum_{n} |\phi^n_{k}(\br_i)|^2 \delta (\omega
- E_{n})$, in the WC phase for bilayer and in liquid phase for a
single layer graphene, see Fig. \ref{FIG:LDOS}.

DC transport measurements is a  less direct alternative  way to
detect WC phase. Depinning effects and other transport anomalies
could be another way to detect suggested WC phase.

In conclusion, we show that the single layer graphene always
remains in a liquid phase.  We also show that the bilayer graphene
is bound to undergo Wigner crystallization at reasonable value of
the Coulomb interaction. We find that in principle there are two
crystallized configurations possible in a bilayer. The in-phase
KED-A state  is stabilized by interlayer hopping. The out-of-phase
WC-B state is stabilized by the Coulomb interactions. LDOS maps,
that can be measured with STM, would reveal charge alternations
between sites and would allow direct imaging of the WC state in
bilayer graphene.

We are grateful to E. Andrei, A. Geim, A. Lichtenstein, M.
Katsnelson, A. Morpurgo, A.C. Neto,  G. Lee, P. Littlewood and H.
Fukuyama for useful discussions. We would like to thank Andrew
Heim for his help to develop the code for the iteration. This work
has been supported by US DOE.

\bibliographystyle{apsrev}
\bibliography{references}

\begin{thebibliography}{14}
\bibitem[{not()}]{note1}
\bibinfo{note}{For simplicity of calculation we introduce an exponential
  cut-off so that potential rapidly drops out at large distances. This
  approximation allows us to keep Coulomb interaction between few sites but
  avoids the difficulty of keeping track for interactions at large distances.
  Since we find a stable Wigner Crystal phase with the potential we consider we
  expect longer range Coulomb terms will only stabilize it more.}
\end{thebibliography}
\end{document}